\documentstyle[aps,prb,twocolumn,psfig]{revtex}
\begin{document}
\twocolumn[\hsize\textwidth\columnwidth\hsize\csname@twocolumnfalse\endcsname 
\title{The Mesoscopic SNS Transistor}
\author{Frank K. Wilhelm$^{1}$, Gerd Sch\"on$^{1}$, and Andrei
D. Zaikin$^{1,2}$} 
\address{$^{1}$ Institut f\"{u}r Theoretische Festk\"{o}rperphysik,
Universit\"{a}t Karlsruhe, D-76128 Karlsruhe, Germany\\
$^{2}$ P.N. Lebedev Physics Institute, Leninskii prospect 53,
117924 Moscow, Russia}
\maketitle
\begin{abstract}
In a mesoscopic superconductor - normal metal - superconductor
(SNS) heterostructure the quasiparticle distribution
can be driven far from equilibrium by a voltage applied across 
the normal metal. This reduces the supercurrent between the
superconducting electrodes, which creates the possibility to use these
SNS junctions as fast switches and transistors.
We describe the system in the framework of the quasiclassical theory
and find good agreement with recent experiments. 
We propose further experimental tests, for instance the
voltage-dependence of the 
current-phase-relation, which includes a transition
 to a $\pi$-junction.
\end{abstract}
]
 Nonequilibrium effects in  superconducting systems~\cite{AS,Noneqsl}
have been gaining new attention due to the increased activities
in the field of mesoscopic electron transport~\cite{Lotsofstuff}.
In contrast to earlier work on nonequilibrium
 superconductivity the new experiments show non-local
 and size-dependent ($d$) effects by reaching
 temperatures below the characteristic Thouless energy 
$E_{\rm Th}={\cal D}/d^2$.  The understanding of this regime is not only of 
fundamental interest, but  also important for nano-electronic
 applications. The  size reduction of electronic devices 
is accompanied by an increase in operation frequency. For instance, in 
the system considered below the latter is limited by $E_{\rm Th}$. 

Recently Pothier {\sl et al.}~\cite{Pothier} probed
 the  quasiparticle properties in
short diffusive wires by coupling tunneling 
 contacts to it. They found that in mesoscopic wires
 the distribution function has a nonequilibrium energy dependence, 
with a double-step structure at the electrochemical potentials of 
both reservoirs. From the smearing of this distribution they further  drew 
conclusions about inelastic relaxation  processes. 
In different setups,  sketched in Figs.\ 1,
Morpurgo {\sl et al.}~\cite{Morpurgo} 
demonstrated that in a superconductor -- normal metal 
heterostructure the nonequilibrium quasiparticle 
distribution due to a  normal current flow in N can be used to tune
 the supercurrent between the superconducting electrodes.
This opens the perspective to use such devices as ultrafast transistors. 

To account for their experimental findings,
Morpurgo {\sl et al.}~\cite{Morpurgo} 
proposed a qualitative  model based on a quasi-equilibrium distribution
function with locally enhanced effective electron temperature. 
This picture is appropriate in the  
limit of strong electron-electron interactions. However, 
 inelastic processes have only a weak effect in the mesoscopic
sample considered here. In fact we find the best performance of
the devices in the opposite limit.

In this article we will describe mesoscopic
superconductor -- normal metal heterostructures as shown in Fig.\ 1 a
and b. For a quantitative analysis
we  use the quasiclassical theory.
It accounts well for the relevant physics:\\
(i) The spectral properties in the normal metal are modified
by the proximity effect due to the presence of the
superconducting electrodes. \\
(ii)  The nonequilibrium quasiparticle distribution
function is found as the solution of a kinetic equation.
If the normal wires between the normal reservoirs 
is shorter than the inelastic scattering
length, the quasiparticle distribution function $f$ has the
nonequilibrium (two-step) form observed in the experiments of Pothier {\sl et
al.}~\cite{Pothier}. \\
(iii) The nonequilibrium distribution influences (reduces) the
supercurrent between the two superconducting contacts through the
normal metal. 

\begin{figure}[htb]
 \centerline{\psfig{figure=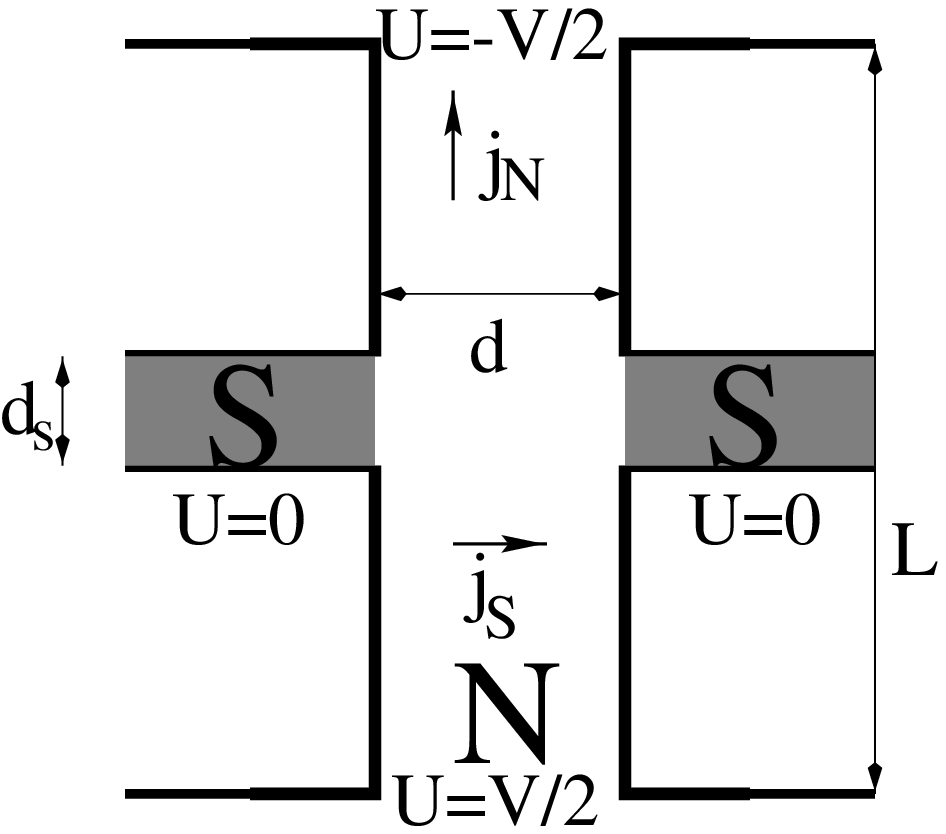,width=40mm}
 \psfig{figure=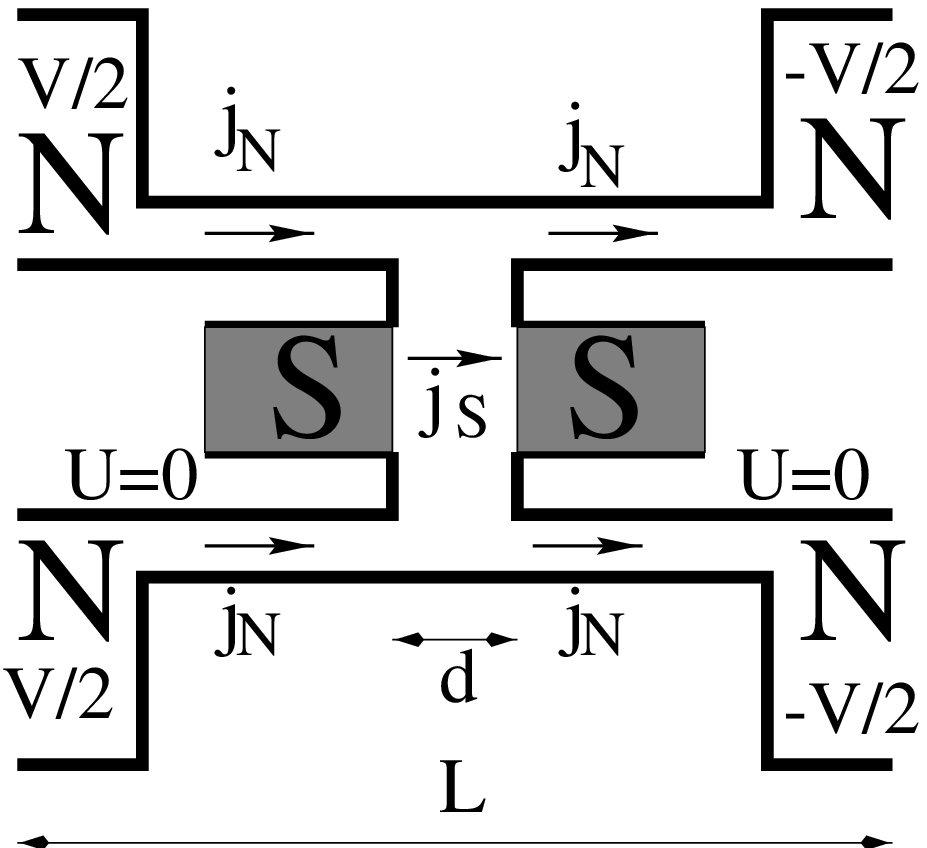,width=40mm}}
\vspace{3mm}
\caption[]{Different realizations of the SNS transistors as used in
Refs.~[\onlinecite{Morpurgo}]. The 
supercurrent is tuned by
(a) a perpendicular or (b) a parallel normal current flow.}
\end{figure}
For best performance as a transistor, in the setup of Fig.\ 1a, 
the width of the superconducting contacts $d_{\rm S}$ should be chosen
narrow compared to the width and length of the normal wire, $d$
and $L$, respectively. The first condition assures that only a small
fraction of the control normal current is diverted through the
superconductors, and accordingly the quasiparticle distribution
function in N is little disturbed by the presence of the
superconducting electrodes. The second condition assures that the
voltage is nearly constant along the superconducting leads, while
the total voltage drop responsible for the nonequilibrium and
reduction of the supercurrent may be large. 
Since the effect relies on a deviation of the distribution function 
from local equilibrium with shifted electro-chemical potential,
the normal wire should have mesoscopic dimensions, i.e.\ the
length $L$ should not exceed the inelastic relaxation length $l_{\rm in}$. 

The presence of the superconducting electrodes induces 
correlations in the normal metal (proximity effect), 
which are responsible for a supercurrent. Their decay length depends
on energy: correlations with energy $\epsilon\gg E_{\rm Th}$ 
decay exponentially, while those within the range of order $E_{\rm Th}$
carry the supercurrent. If, in a nonequilibrium situation, these
states are occupied and in this way blocked for superconducting
correlations,  the superconductivity is weakened and the 
supercurrent reduced. This suppression mechanism will be described
in the following, based on the real-time formalism of quasiclassical
Green-Keldysh functions in the diffusive limit \cite{LO,AS,VZK}. 

In the first step we describe the proximity effect in the normal metal
by analyzing Usadel's equations.
The standard parameterization of normal 
and anomalous retarded Green functions $G^{\rm R}=\cosh\alpha $ and 
$F^{\rm R}=\sinh\alpha e^{i\chi}$ allows us to 
write these equations, for the normal metal region between the superconducting
electrodes, in the form
\begin{eqnarray}
{\cal D}  \partial^2_x\alpha &=&-2i\epsilon\sinh\alpha -({\cal D}/2)
\left(\partial_x\chi \right)^2\sinh2\alpha \nonumber\\
\partial_xj_\epsilon&=&0\;\;\; \mbox{,} \;\;\;
j_\epsilon=(\partial_x\chi )\sinh^2\alpha \; . 
\label{retard}
\end{eqnarray}
Here, ${\cal D}$ is the diffusion coefficient and $x$ the coordinate
normal to the NS interfaces. We also introduced the energy-dependent
`spectral current' $j_\epsilon$. For simplicity we ignored here a
dependence of the spectral quantities 
on the coordinate $y$ parallel to the interfaces.
This dependence should  merely lead to a quantitative modification
of our results and conclusions.
In the realistic limit $\Delta\gg E_{\rm Th}$
and for transparent metallic interfaces the boundary conditions at these
interfaces $x=\pm d/2$ read (see [\onlinecite{WSZ}]
and Refs. therein)
\begin{equation}
  \alpha (\pm d/2)=-i\pi/2\;\;\; \mbox{,} \;\;\;
  \chi (\pm d/2)=\pm\phi/2.
\label{boun}
\end{equation}
\begin{figure}
 \centerline{\psfig{figure=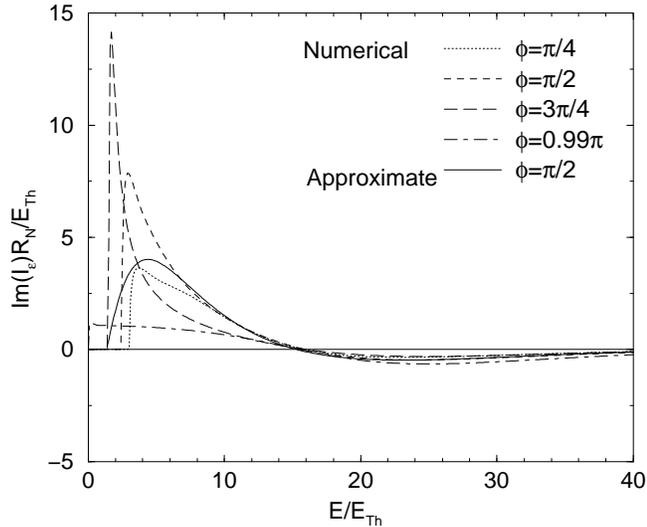,width=85mm}}
\caption{The spectral current $\hbox{Im}(j_\epsilon)$ as a function 
of energy for different values of the phase difference $\phi$.}
\end{figure}

In two limits we find analytic solutions of 
Eqs. (\ref{retard}, \ref{boun}).  For low energies
 $\epsilon\ll E_{\rm Th}$ we obtain
\begin{eqnarray*}
  \alpha &\simeq &-i\pi/2+(\epsilon/E_{\rm Th})\, a(\phi)+O(\epsilon^3) \\
\chi &\simeq &\phi\, x/d-(\epsilon/E_{\rm
Th})^2\,b(\phi)+O(\epsilon^3) \; ,
\end{eqnarray*}
where $a(\phi)$ and $b (\phi )$ are real-valued functions 
(omitted for brevity), while for high energies 
$\epsilon \gg E_{\rm Th}$ 
we obtain \cite{WSZ}
\begin{eqnarray}
  F^{\rm R}&=&F_0(x-d/2)e^{i\phi/2}+F_0(d/2-x)e^{-i\phi/2}
\label{anaj}\\
F_0(s)&=&4q\frac{1+q^2}{(1-q^2)^2}\;\; \mbox{,} \;\;
q(s)=i(\sqrt{2}-1)e^{-s\sqrt{-2i\epsilon/E_{\rm Th}}} \nonumber\\
\hbox{Im}(j_\epsilon)&=&64\sin\phi \, \hbox{Im}
\left(\left.\frac{(1+q^2)(1+6q^2+q^4)}{(1-q^2)^5}qq^\prime
\right)\right|_{s=L/2} \; .\nonumber
\end{eqnarray}
In addition we have studied the problem numerically.
Combining our results in Fig.\ 2  we observe the
following features of the spectral current $\hbox{Im}(j_\epsilon)$.
It is an odd function of $\epsilon$ and
shows a proximity induced mini-gap~\cite{GolKupr}, 
 $\epsilon_{\rm g} \simeq 3.2 E_{\rm Th}$ at $\phi =0$, 
below which  $\hbox{Im}(j_\epsilon)=0$. 
This gap decreases with increasing
$\phi$ \cite{Charlat} and vanishes at $\phi =\pi$. 
At energies directly above the gap, $\hbox{Im}(j_\epsilon)$
increases  sharply, but rapidly decreases at higher $\epsilon$. 
At large energies, it changes sign 
and oscillates around zero with exponentially 
decaying amplitude.

 Next we determine the nonequilibrium quasiparticle
distribution in the normal metal
between the reservoirs, which are at different
electrochemical potentials $\pm eV/2$. In a mesoscopic length wire 
$L \ll l_{\rm in}$ in the diffusive limit the 
distribution  function obeys the kinetic equation
\begin{equation}
        \partial^2_yf=0 \; .
\label{kin}
\end{equation}
In the absence of superconducting contacts
its solution 
\begin{eqnarray}
f(\epsilon,y)&=&\left(1/2-y/L\right)f^{\rm eq}\left(\epsilon +eV/2\right)\nonumber\\
&&+\left(1/2+y/L\right)f^{\rm eq}\left(\epsilon-eV/2\right)
\label{flres}
\end{eqnarray}
has  two temperature-rounded steps
 at the  electrochemical potentials of both
reservoirs. 
The step heights depend on the position along the wire;
in this way the distribution function interpolates linearly 
between the boundary conditions at $y=\pm L/2$. 
This functional dependence had been
detected in the experiments of Pothier {\sl et al.}~\cite{Pothier}.

 Although the distribution function definitely
has not a thermal form, a local electrochemical potential of the normal metal, 
$\mu(y)$, and an effective electron temperature
can be defined by the corresponding moments of the distribution function.
For instance $\mu(y)$ follows from $ \int_{-\infty}^\infty d\epsilon 
[f(\epsilon,y) - f^{\rm eq}(\epsilon - \mu(y))] = 0$, where 
$f^{\rm eq}$ denotes the Fermi function.
In the following we will consider the situation where the
 electrochemical potential of the superconductors coincides with the
local value  of the normal metal, which guarantees
that there is no net current out of the normal metal into the superconductors. 
Since we further have chosen the size of the 
superconducting contacts, $d_{\rm S}$, small compared to the 
width and length, $d$ and $L$, of the normal wire, the
distribution function in N is little disturbed by the contacts.

On the other hand, the quasiparticle distribution function
influences the superconducting correlations induced in the normal
metal. The supercurrent through the heterostructure is given by
\begin{equation}
I_{\rm S}= \frac{d}{2R_d}
\int_{-\infty}^\infty d\epsilon\; [1-2f(\epsilon,y_{\rm S})]
\, \hbox{Im}(j_\epsilon) \; ,
\label{eq:supc}
\end{equation}
where $1/R_d=2e^2N_0{\cal D}S/d$, $S$ is the junction area, and
$y_{\rm S}$ denotes the position of the superconducting electrodes.
It is important to note that the energy is
measured relative to the electrochemical potential of the 
superconductors, which we have chosen to coincide with the local value
of the normal metal, $\mu(y)$.
Due to the odd symmetry of $\hbox{Im}(j_\epsilon)$, only the odd
component of a nonequilibrium quasiparticle distribution
 modifies $I_{\rm S}$. 
Accordingly the first term in the integral (\ref{eq:supc}) can 
be written as\cite{LO} $-f(\epsilon,y_{\rm S}) + f(-\epsilon,y_{\rm S})$, which
displays that an
excess number of electron-like  or of hole-like excitations 
have the same effect on the supercurrent. 

The largest effect is found when the superconducting
electrodes are placed symmetrically between the two normal
reservoirs (i.e. at $y=0$, see eq. (\ref{flres})). 
The resulting modification of the supercurrent across the SNS junction is
 presented in Fig.\ 3 as a function
of the  voltage $V$ across the normal metal.

\begin{figure}
 \centerline{\psfig{figure=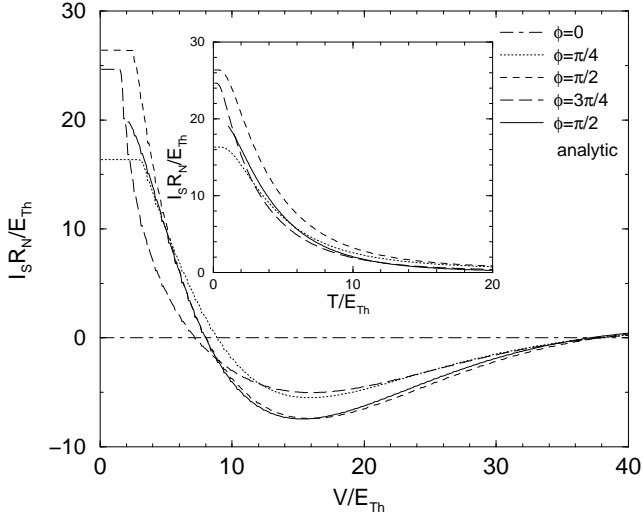,width=85mm}}
\caption{The supercurrent as function  of  control
voltage (at $T=0$) and temperature
(at $V=0$) for various values of $\phi$.}
\end{figure}

At low temperatures $T \ll eV$, $f(\epsilon,0)$ 
deviates from the equilibrium value only in the window
 $-eV/2 <\epsilon < eV/2$.
Since the spectral current Im$j_\epsilon$ vanishes for 
$\epsilon <\epsilon_{\rm g}$
there is no modification of 
of the supercurrent for small voltages $eV < 2\epsilon_{\rm g}$.
On the other hand, for $eV>2\epsilon_{\rm g}$, 
extra quasiparticle and hole states with energies 
$|\epsilon |$ below $eV/2$ are occupied and the supercurrent is diminished.
Since this energy window
increases with increasing $V$ the supercurrent decays rapidly
with voltage (cf. Fig. 3). This is exactly what has been observed
in the experiments \cite{Morpurgo}. Furthermore, at still larger
voltage $eV \gtrsim 10 E_{\rm Th}$ the supercurrent changes sign
since the integral in (\ref{eq:supc}) is dominated by the energy
interval where Im$j_\epsilon <0$ (Fig. 2). 
We thus find a transition to a
so-called $\pi$-junction \cite{Bul}, controlled by 
nonequilibrium effects. This effect is rather 
pronounced,  the critical current of the $\pi$-junction
is approximately  30\% of $I_{\rm c}$ at $T=eV=0$.

For high voltages or temperatures
 $eV, T \gg \epsilon_{\rm g}$ we find 
 $I_{\rm S}=I_{\rm c}\sin{\phi}$ with critical current 
\begin{eqnarray}
  I_{\rm c}R_d&=&\frac{64\pi}{3+2\sqrt{2}}
e^{-\sqrt{\Omega_V/E_{\rm Th}}}\cos\left(\frac{eV}{2\sqrt{\Omega_VE_{\rm Th}}}+
\varphi_0\right)\nonumber\times\\
&&\times
\cases{eV&for $eV\gg \epsilon_{\rm g}$, $T\ll \epsilon_{\rm g}$\cr
T\left(\frac{{(2\pi T)^2+e^2V^2}}{E^2_{\rm Th}}\right)^{\frac14}&for 
$T\gg \epsilon_{\rm g}$\cr}\label{exp} \; .
\end{eqnarray}
Here $\Omega_V = \pi T+\sqrt{(\pi T)^2+(eV/2)^2}\label{exp2}$ and 
$$
\varphi_0 = \cases{\pi/2&for $eV\gg \epsilon_{\rm g}$, $T\ll \epsilon_{\rm g}$\cr
\frac{1}{2}\tan^{-1}\left(\frac{eV}{2\pi T}\right)&for $T\gg 
\epsilon_{\rm g}$\cr}.
$$

Nonequilibrium effects also influence the current-phase-relation
$I_{\rm S}(\phi)$. The rich variety of different 
curves at different voltages is  displayed
in Fig. 4. In contrast a quasi-equilibrium theory would always predict
a functional dependence as shown in the inset of the figure.

\begin{figure}
\centerline{\psfig{figure=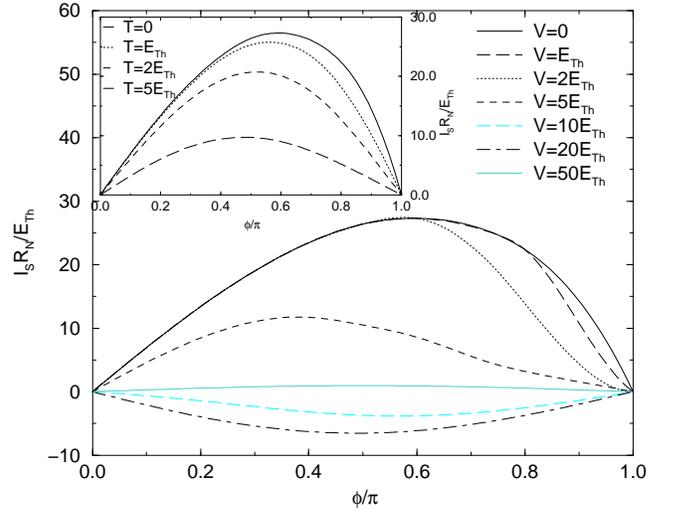,width=85mm}}
\caption{The supercurrent-phase relation at different temperatures and
control voltages.}
\end{figure}

Since the supercurrent decays exponentially as a function of both $T$ 
and $V$, one might try to describe the system properties
 by a quasi-equilibrium theory with 
effective $V$-dependent temperature $T^*$. 
Exactly this strategy was adopted in
 Ref. [\onlinecite{Morpurgo}]
 with $T^* = \sqrt{T^2+\gamma^2(eV)^2}$. The 
the best fit to the experimental data was obtained for $\gamma\simeq
 6K(mV)^{-1}$.  
On the other hand, as  demonstrated above, 
the analogy between temperature and voltage is incomplete and even
misleading. The distribution function in the normal metal  
 in general {\it cannot} be described by the Fermi function with
effective temperature $T^*$. One of the most striking consequences,
 the transition to a $\pi$-junction, is not obtained in the
 quasi-equilibrium description. 
Apart from this qualitative difference,  the dependence of
$I_{\rm c}(V,T)$ (\ref{exp}) deviates from 
that suggested in Ref. [\onlinecite{Morpurgo}]. While we find
 a fairly good agreement between our results and the data \cite{Morpurgo}
 at the temperatures of the experiment, we cannot summarize
 our results in general by an effective temperature model. It would be
desirable to expand the experiments to a parameter range where
 the difference between the two descriptions become more pronounced.

The configuration of Fig. 1a is but one realization
of a mesoscopic system where the supercurrent can be controlled by
an externally applied voltage. Another realization,
also studied in Ref. [\onlinecite{Morpurgo}] is depicted in Fig. 1b.
In this case the distribution function in the N-layer between two
superconductors is driven out of equilibrium by the normal 
current flowing parallel to the supercurrent. Provided $d \ll L$
there is practically no voltage drop
across the SNS junction and only dc Josephson effect can be considered.  
Since the distribution function  
has  the same form (\ref{flres}) as before,
the previous results for $I_{\rm S}(V)$ apply also for the structure of Fig.\ 1b.
Again good agreement with the experimental findings \cite{Morpurgo}
is observed.

Yet another system with a voltage controlled supercurrent
 was studied by Volkov
\cite{Volkov1}. He considered a SINIS system where the
normal metal was thin and  separated by 
low transparency barriers from the superconductors.
In this case the superconductors and the normal metal
are in equilibrium, except that
their electrostatic potentials are shifted relative to each other,
with the total voltage drop across the  barriers. 
Although this nonequilibrium situation is very different from the one
discussed here, he
 also finds a  voltage-dependent supercurrent
reduction as well as a transition to a $\pi$-junction.

If $L$ is not large compared to $d_{\rm S}$ and $d$, 
the  conversion between super- and normal currents
in the junction area cannot be neglected and the
theoretical analysis has to be extended. For instance
 the distribution functions depend both on $x$ and $y$.
A  thorough analysis of  the microscopic theory~\cite{LO,AS} 
reveals further that the expression for the current (\ref{eq:supc})
has to be extended, and that
odd and even component (in energy) of the distribution function, 
 $f_{\rm L}$ and $f_{\rm T}$, obey two  coupled diffusion  equations,
but with different, position-dependent effective diffusion coefficients. 
They are coupled by terms of the form 
$\hbox{Im}j_\epsilon\cdot\nabla f_{\rm L/T}$
In this case the suppression of $f_{\rm T}$ influences $f_{\rm L}$ 
and  weakens the performance of the transistor, although qualitatively
the physical situation remains unchanged. 

We now turn to the important practical question:  how 
efficient is this device as a transistor? 
The control and signal voltage, $V$ and
$V_{\rm S}=I_{\rm c}R_d$, are both of the order of the Thouless
energy $E_{\rm Th}$. Thus, no voltage gain is obtained.
However, the power amplification is proportional to the ratio of the
relevant resistors $R_L/R_d$. Here  $R_L$ is the
resistance of the normal metal of length $L$, while  $R_d$, introduced
in  (\ref{eq:supc}), is
the resistance between the superconducting electrodes in a situation
where $I_{\rm c}$ is low and this transport is dissipative as
well. Since both are governed by the same material-dependent
conductivity the ratio depends on the relevant lengths, 
$R_L/R_d \propto {L d_{\rm S}}/{d^2}$.
Hence, by choosing a sufficiently long control line, 
 $L\gg d,d_{\rm S}$, a power amplification can be achieved. The
limitation in operation frequency in the mesoscopic regime is also
provided by the Thouless energy.

In summary, we have presented a microscopic description of  
nonequilibrium electronic properties of mesoscopic SNS heterostructures.
The distribution function in the normal metal can be driven far from
equilibrium by a voltage applied at a distance $\sim L$
from the junction. This distance is
limited only by the inelastic relaxation length $l_{\rm in}$.
We analyzed how the supercurrent across the sample is reduced by this 
control voltage. The strongest reduction and, hence, best performance of 
the device is found in a mesoscopic situation,
when the distribution function deviates significantly from a local
equilibrium form. We established the connection to experiments and 
suggested further  tests. 
The possibility to control the supercurrent by an external voltage
 allows several technical applications,
for instance the use as a high-frequency transistor with power gain
proportional to the ratio between  length and width of the normal wire.

We want to express our respect to the late Albert Schmid, who
pioneered  the theory on which this work is based. 
We acknowledge further discussions with  R.\ Raimondi, A.\ Morpurgo,
J.J.A.\ Baaselmans and  H.\ Pothier. This work was  supported by
the DFG through SFB 195 and the Graduiertenkolleg 
``Kollektive Ph\"anomene im Festk\"orper'', and  INTAS Grant 93-790-ext.

\end{document}